\RequirePackage{fix-cm}
\documentclass[twocolumn]{article}
\usepackage{hyperref}
\usepackage[left=2.0cm,right=2.0cm,top=2.5cm,bottom=2.5cm]{geometry}
\usepackage{color}
\usepackage{graphicx}
\usepackage{amsmath,amssymb}
\usepackage{caption}
\usepackage{subcaption}
\title{Unilateral vibration transmission in mechanical systems with bilinear coupling
}

\author{Ali Kogani\footnotemark[1] 
\and Behrooz Yousefzadeh\footnotemark[2]}

\begin{document}
\maketitle
\renewcommand{\thefootnote}{\fnsymbol{footnote}}
\footnotetext[1]{Department of Mechanical, Industrial \& Aerospace Engineering, Concordia University, Montreal, QC, H3G1M8, Canada(\href{mailto:ali.kogani@mail.concordia.ca}{ali.kogani@mail.concordia.ca})
}
\footnotetext[2]{Department of Mechanical, Industrial \& Aerospace Engineering, Concordia University, Montreal, QC, H3G1M8, Canada(\href{mailto:behrooz.yousefzadeh@concordia.ca}{behrooz.yousefzadeh@concordia.ca})}

\begin{abstract}
Unilateral transmission refers to the scenario in which the waves transmitted through a system remain in pure tension or pure compression. This transmission phenomenon may occur in systems that exhibit different effective elasticity in compression and tension; {\it i.e.} bilinear elasticity. We present a computational investigation of unilateral transmission in the steady-state response of harmonically driven mechanical systems with bilinear coupling. Starting with two bilinearly coupled oscillators, we find that breaking the mirror symmetry of the system, in either elastic or inertial properties, facilitates unilateral transmission by allowing it to occur near a primary resonance. This asymmetry also enables nonreciprocal transmission to occur. We then investigate the nonreciprocal dynamics of the system, including linear stability analysis, with a focus on unilateral transmission. We also extend our discussion to a bilinear periodic structure, for which we investigate the influence of the number of units and energy dissipation on unilateral transmission. We report on the existence of stable nonreciprocal unilateral transmission near primary and internal resonances of the system, as well as other nonreciprocal features such as period-doubled and quasiperiodic response characteristics.

\end{abstract}


\section{Introduction}
\label{intro}
It is generally preferred in engineering to design vibrating machines and devices to operate in the linear regime, where their behavior is more easily predictable. Nonlinearity, within this framework, is often associated with unwanted complexity. However, nonlinearity is unavoidable in several engineering systems such as systems with friction joints or clearances~\cite{gaul_role_2001, yuan_friction_2024}. On the other hand, nonlinearity can be introduced intentionally to a vibrating system with the goal of achieving new response characteristics~\cite{vakakis_intentional,rajalingham_influence_2003, wagg2012exploiting}. There is therefore a need to understand the nonlinear behavior of vibrating systems from both theoretical and practical perspectives.  

Nonlinearity can originate from the material properties of a system (material nonlinearity)~\cite{dickie1972geometric}, from large deformations (geometrical nonlinearity)~\cite{ghuku2017review} or from the interaction force between multiple sub-systems such as friction in joints (contact nonlinearity)~\cite{gaul_role_2001}. In this work, we focus on a particular type of nonlinearity known as bilinear elasticity or bilinear stiffness. 

Bilinear stiffness refers to a system characteristic where the effective elasticity changes at a certain degree of elastic deformation, exhibiting one stiffness value up to a threshold called the offset, and a different stiffness value beyond that threshold. This unique nonlinearity can be found in systems such as a cracked beam~\cite{chance_simplified_1994}, in vibro-impact drilling systems~\cite{liu_analysis_2018}, or in systems with intermittent contact~\cite{saito_node_2010, zucca_bi-linear_2014}. Systems featuring bilinear elasticity are known for their non-smooth behavior (sudden change between two states), leading to intricate nonlinear dynamic response characteristics~\cite{bernardo_piecewise-smooth_2008}. Vibration systems with bilinear stiffness may exhibit unique dynamic behavior not readily found in other types of nonlinearity. Therefore, it is important to understand the associated dynamics and vibration transmission characteristics of such systems.

The study of bilinear stiffness and its impact on the vibration characteristics of mechanical systems has a long history~\cite{SHAW1983129, iwan_steady-state_1965, thompson_subharmonic_1983, wong_periodic_1991}. Extensive research has explored various aspects of vibration characteristics influenced by bilinear stiffness and damping~\cite{ahmed_equivalent_1992, narimani_frequency_2004, jiang_grazing-induced_2017,shi_vibration_2019, fontanela_nonlinear_2021, yastrebov_wave_2022}. In this work, we focus on a response characteristic that can be triggered effectively by bilinear stiffness: unilateral transmission.

Unilateral transmission is a response characteristic where the transmitted wave remains purely in tension or compression; {\it i.e.}, the corresponding deformations from the static equilibrium position remain consistently positive or negative. Lu and Norris~\cite{lu_unilateral_2021} were the first, to the best of our knowledge, to demonstrate unilateral transmission using a single bilinear stiffness connecting two waveguides. They derived equations that describe the relationship between the parameters of the bilinear spring and the conditions necessary for achieving unilateral transmission. Despite the intriguing nature of unilateral transmission, this phenomenon has not been explored in details in the context of vibration dynamics Our main goal is to investigate unilateral transmission in the context of vibration transmission in mechanical systems, with a focus on understanding the influence of different system parameters on this phenomenon.

We investigate unilateral transmission using an archetypal system of two coupled oscillators. Each oscillator may be viewed as the amplitude equation for linear waves propagating in a waveguide. Bilinearity is introduced through the coupling spring; this is the only source of nonlinearity in this work. We focus exclusively on steady-state vibration transmission in response to an external harmonic force. We find that breaking the mirror symmetry of the system facilitates unilateral transmission by allowing it to occur near a primary resonance. 
 
A consequence of breaking the mirror symmetry of the system is that it enables nonreciprocal dynamics~\cite{nassar_nonreciprocity_2020}. Nonreciprocity in systems with bilinear coupling has already been reported, primarily with a focus on attaining a significant difference in transmitted energies in opposite directions~\cite{lu_non-reciprocal_2020, lu2021nonreciprocal,wallen2018static,yastrebov,dang}. Here, we investigate nonreciprocity in the context of unilateral transmission. We show that nonreciprocal unilateral transmission may occur either in one transmission direction or both. In addition, we report on other nonreciprocal features of the response such as direction-dependent appearance of higher harmonics, periodic-doubled response or quasiperiodic dynamics, some of which are not presented often in the literature. 

Inspired by recent advances in the manufacturing of periodic lattice materials with tunable dynamic properties \cite{wang_tunable_2020}, we extend our analysis to a periodic lattice made from repetition of bilinearly coupled units. We investigate the wave propagation characteristics of a periodic material with a bilinear unit cell, with a focus on analyzing unilateral transmission. We then conduct a parametric study to investigate the influence of the number of units and of energy dissipation on unilateral transmission in the periodic lattice.

Section~2 presents the problem formulation and solution methodology. We discuss unilateral transmission for bilinearly coupled systems in Section~3. We extend the analysis to a periodic structure made from bilinearly coupled units in Section~4. We review our main findings and conclude in Section~5.

\section{Problem setup}
\label{sec:1}

\subsection{Governing equations}
\label{EOM}

\begin{figure}

  \includegraphics[width=.35\textwidth]{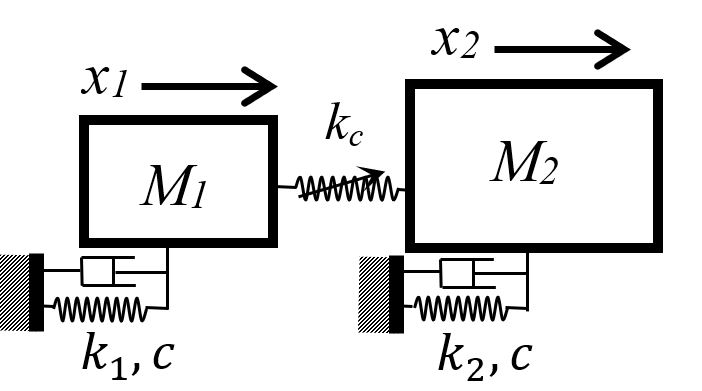}
  \centering
\caption{Schematic of the 2DoF system consists of two linear damped oscillators coupled with a bilinear spring. }
\label{fig:1}      
\end{figure}

Fig.~\ref{fig:1} shows the schematics of the two-degree-of-freedom (2DoF) vibration system we study in this section. The system consists of two masses, $M_1$ and $M_2 = \mu M_1$, connected by a bilinear spring $k_c$. The two masses are anchored to the ground by linear springs $k_1$ and $k_2$. Energy loss is accounted for by identical linear viscous dampers with a constant $c$, connecting each mass to the ground. The system is subject to external harmonic excitation with amplitude $F$ and frequency $\omega_f$, which acts on one of the two masses depending on the transmission scenario. The symmetry of the setup can be controlled by either the mass ratio $\mu = \frac{M_2}{M_1}$ or the ratio of the grounding springs $r = \frac{k_2}{k_1}$.

As outlined in Appendix A, the equations of motion for the system shown in Fig.~\ref{fig:1} can be written using non-dimensional parameters as 
\begin{equation}
\label{govern}
\begin{aligned}
\ddot{x}_1+K_c(x_1-x_2)+x_1+2\zeta\dot{x}_1=F_1\cos{\omega_ft} \\
\mu\ddot{x}_2+K_c(x_2-x_1)+rx_2+2\zeta\dot{x}_2=F_2\cos{\omega_ft}
\end{aligned}
\end{equation}
where 
\begin{equation}
\label{bilinear}
  K_c =
    \begin{cases}
      \beta k, & x_2-x_1>0\\
      k, & x_2-x_1<0\\
    \end{cases}       
\end{equation}

The parameter $K_c$ represents the bilinear stiffness, which results in a nonsmooth (piecewise smooth) force in the system. This is the origin of nonlinearity in our system. We define $\beta$ as the bilinear ratio. The coupling stiffness (therefore the coupling force) is considered softening when $\beta<1$ and hardening when $\beta>1$. The damping ratio is denoted by $\zeta$ and the forcing frequency by $\omega_f$.

In the proceeding sections, we study the end-to-end vibration transmission characteristics of Eq.~(\ref{govern}). For this purpose, we define (i) the {\it forward} configuration with $F_1=P$ and $F_{2}=0$, in which the output is the steady-state displacement of the last (right-most) mass, $x_{2}^F(t)$; (ii) the {\it backward} configuration with $F_1=0$ and $F_{2}=P$, in which the output is the steady-state displacement of the first (left-most) mass, $x_1^B(t)$.
To quantify the degree of nonreciprocity of the response, we define the reciprocity bias, $R$, as
\begin{equation}
\label{R}
R = \frac{1}{T} \int_0^T \left(x_2(t)^F - x_1(t)^B\right)^2 \, dt
\end{equation}
where $T=2\pi/\omega_f$ is the forcing period. We have $R=0$ if and only if the response is reciprocal.

\subsection{Solution methodology}
The bilinear stiffness represents a strong nonlinearity. The nonsmooth nature of this nonlinear force poses challenges for the analysis of the system. It is possible to obtain analytical expressions in the frequency-preserving (weakly nonlinear) operating regime of the system~\cite{marino_displacement_2019}. This approach is cumbersome for systems with many degrees of freedom, however. Direct numerical computation of the response of the system is also time-consuming for performing a parametric study. 

Given that our focus is exclusively on the steady-state response of the system to harmonic excitation, we will use continuation techniques to compute the response of the bilinear system. It is possible to capture the nonsmooth nature of the bilinear force with no approximation, but the associated cost is increasing the computation time. Instead, we approximate the bilinear stiffness by the following smooth function: 
\begin{equation}
\label{regularization}
  K_c =\frac{k(\beta-1)}{\pi}\arctan{(B(x_2-x_1))}+\frac{k(\beta+1)}{2}     
\end{equation}
The accuracy of this approximation ({\it regularization}) depends on the value of the regularization parameter, $B$. Higher values of $B$ result in a sharper transition around the transition point between the two values of the stiffness, $k$ and $\beta k$, at the cost of making the equations numerically stiffer. We have chosen $B=1000$ for all the results presented in this work. This will be validated by comparing the results with those from an event-driven numerical integration scheme. When the transition between the two values of stiffness occurs at zero deformation, we refer to this case as bilinear stiffness with no offset. The steady-state response of a bilinear system with no offset is independent of the value of the forcing amplitude \cite{shi_vibration_2019}. For the low to moderate values of forcing amplitude that we consider in this work, the response is indeed independent of the forcing amplitude -- see Appendix B. When the offset is not zero, the response below a certain threshold is linear (amplitude independent) until the bilinearity is triggered; see \cite{kogani_nonreciprocal_2022} for an example. 

We use the software package {\sc coco}~\cite{coco} to compute the steady-state response of the system under harmonic excitation as a family of periodic orbits with period $T=2\pi/\omega_f$ that satisfy a suitable boundary-value problem~\cite{doedel_lecture_2007}. This is particularly suitable for our system because we anticipate the response to be anharmonic within the parameter ranges that we explore. The (local) stability of the response is determined by the Floquet multipliers associated with each periodic orbit.

Because the restoring force in the bilinear spring, $f(x_2-x_1)=K_c(x_2-x_1)$, is not symmetric with respect to its deformation, $f(d)\ne f(-d)$, there will be a drift (DC term) in the steady-state response of the system; {\it i.e.} the oscillations may not be centered around the static equilibrium points of the two masses. To account for this, we define the DC value $C_i$ around which each mass $i$ oscillates as
\begin{equation}
\label{DC}
C_i=\frac{\textnormal{Max}\left(x_i\left(t\right)\right)+\textnormal{Min}\left(x_i\left(t\right)\right)}{2}\      
\end{equation}
We define the amplitude of the oscillation or the AC value for each mass as
\begin{equation}
\label{AC}
A_i=\frac{\textnormal{Max}\left(x_i\left(t\right)\right)-\textnormal{Min}\left(x_i\left(t\right)\right)}{2}     
\end{equation}
which represents the maximum deflection of each mass from its static equilibrium position. 

\begin{figure}
\begin{subfigure}[b]{.24\textwidth}
  \includegraphics[width=\textwidth]{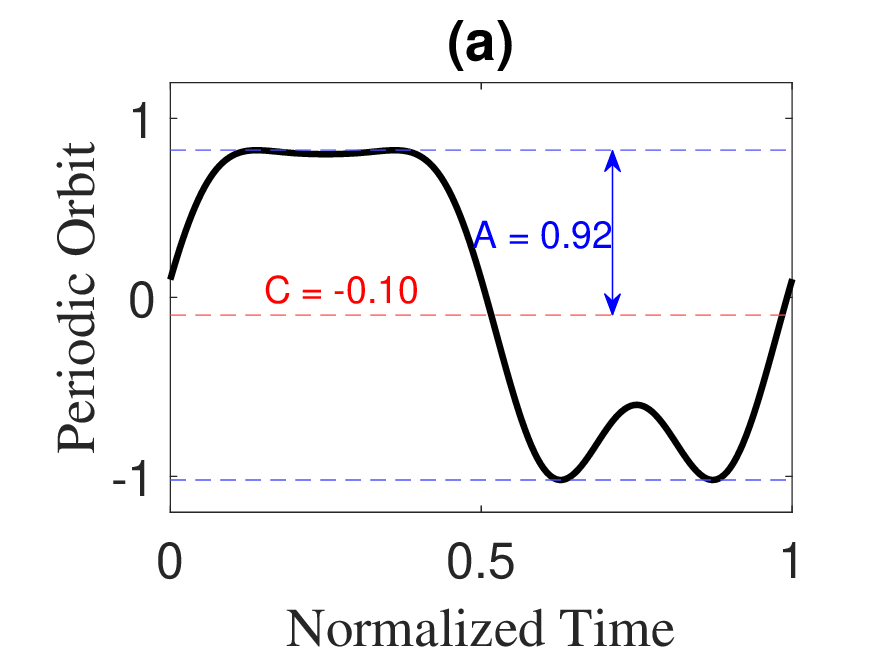}
\end{subfigure}
\hfill
\begin{subfigure}[b]{.24\textwidth}
      \includegraphics[width=\textwidth]{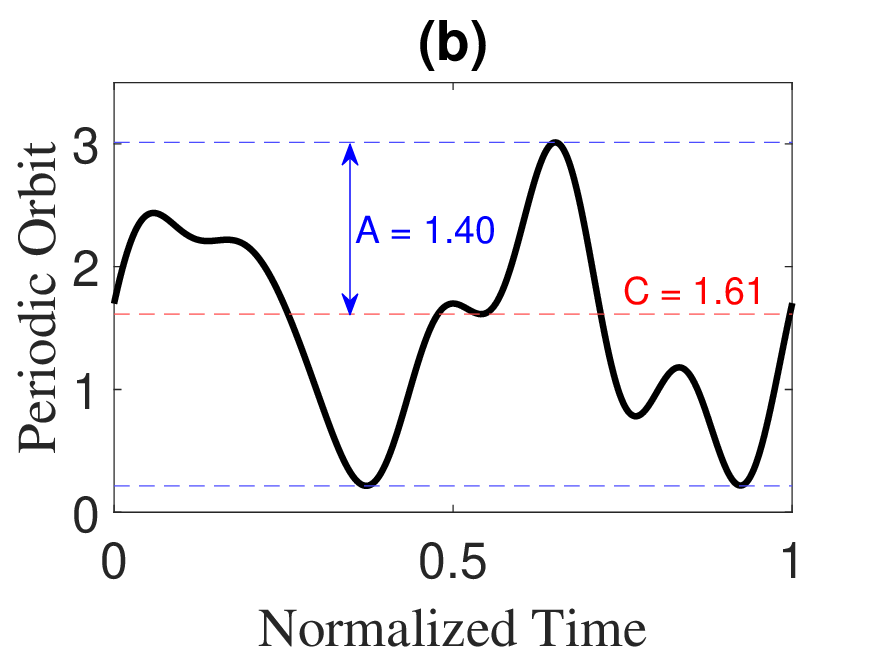}
\end{subfigure}

    \caption{Periodic orbits represented by $A$ and $C$ (a) non-unilateral (b) unilateral}
    \label{fig:2}
\end{figure}

We note that the choice of amplitude parameters is not unique. Our goal here is to use parameters that help determine when a response is unilateral. Fig.~\ref{fig:2} shows how parameters $A$ and $C$ in Eqs.~(\ref{DC}-\ref{AC}) represent two examples of anharmonic periodic solutions. Panel~(a) shows a periodic response that is not unilateral; {\it i.e.}, $|A|>|C|$. Panel~(b) shows a unilateral response for which $|A|<|C|$. Of course, these norms also work well when the response is harmonic. We can therefore use the ratio of $C$ to $A$ in the output displacement to determine whether the transmitted vibration is unilateral.

For unilateral transmission to occur, the output must exhibit a minimum value greater than zero or a maximum value less than zero. In other words, the transmitted vibration displacement must be exclusively positive or negative. Using the norms $A$ and $C$ from Eqs.~(\ref{DC}-\ref{AC}), we define the unilateral ratio as
\begin{equation}
\label{Ru}
R_u=\frac{|C_i|}{|A_i|}    
\end{equation}
where $i=2$ for the forward configuration and $i=1$ for the backward configuration. Thus, $R_u>1$ represents unilateral transmission and $R_u=1$ corresponds to the onset of unilateral transmission.

\section{Bilinearly coupled oscillators}
\label{unilateral}

\subsection{Unilateral transmission in a symmetric system}
\label{sym}
We first focus on unilateral transmission in a 2DoF system with mirror symmetry (identical oscillators), characterized by $\mu=1=r$. There is no difference between the left-to-right and right-to-left transmission characteristics in this case and the response of the system is (trivially) reciprocal by virtue of the mirror symmetry. 
We consider the case of softening bilinearity, $\beta<1$, in which the system is stiffer in compression than in tension. This type of bilinear force can be found in coupled structures assembled with bolted joints~\cite{joint}.

\begin{figure*}
\begin{subfigure}{1\textwidth}
  \includegraphics[width=\textwidth]{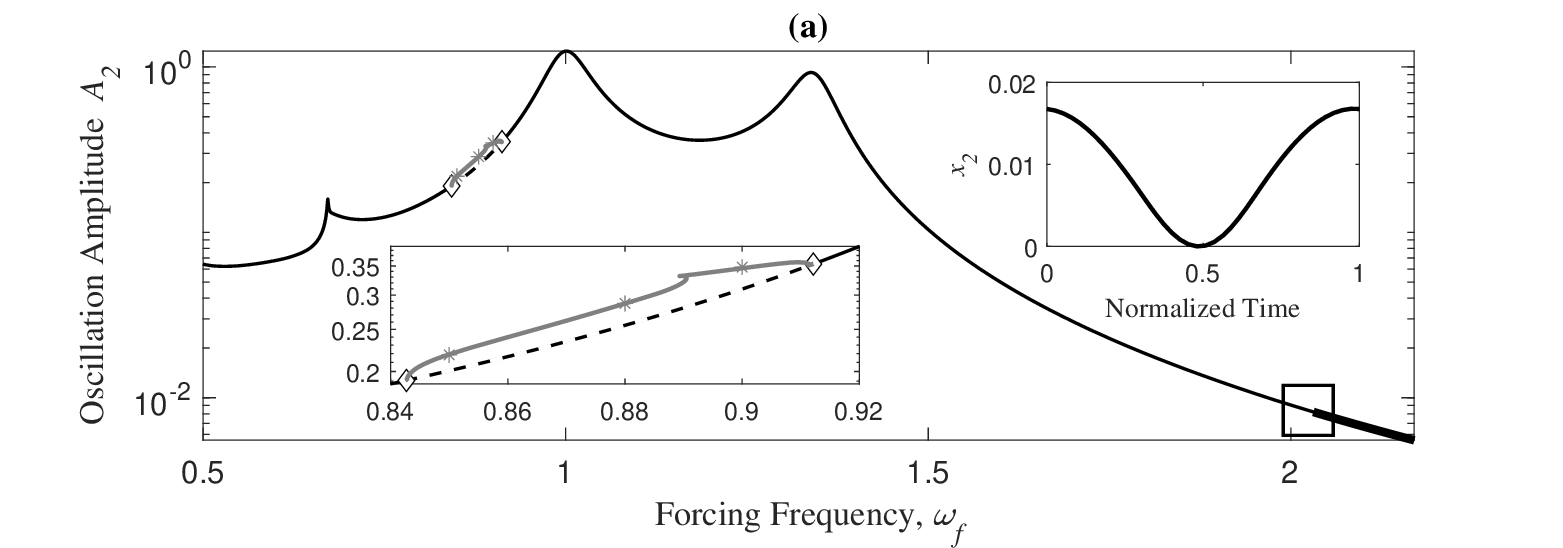}
\end{subfigure}

\begin{subfigure}{1\textwidth}
\includegraphics[width=1\textwidth]{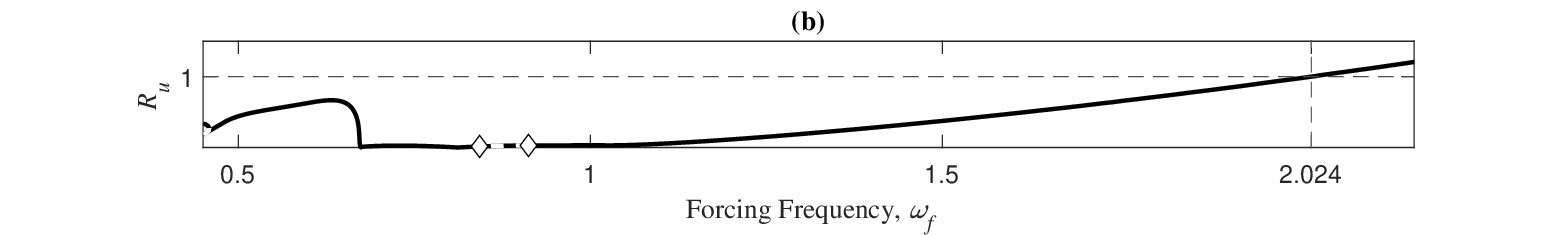}
\end{subfigure}
    \caption{
(a)~Frequency response of the symmetric 2DoF ($r=\mu=1$) system for the forward configuration. Thick lines indicate response that exhibits unilateral transmission and dashed lines indicate unstable response. Diamond markers indicate period-doubling bifurcation points. The upper right inset shows the unilateral time response at the frequency highlighted by the black square.
(b)~Unilateral transmission $R_u$ as a function of forcing frequency.
}
\label{fig:3}
\end{figure*}

Fig.~\ref{fig:3}~(a) shows the frequency response function of the output of the system for $P=0.15, \zeta=0.03, k=1, \beta=0.1$, and $r=\mu=1$. The unilateral ratio, $R_u$, is monitored to find regions with unilateral transmission; these regions are indicated by thicker lines. The top right inset shows the steady-state output displacement of the system during one period ($2\pi/\omega_f$) at the onset of unilateral transmission, $R_u=1$, near $\omega_f\approx2.03$. The displacement is always positive in sign (purely in extension) and does not cross the zero line (static equilibrium position). The unstable range below the first primary resonance ($0.84 < \omega_f < 0.92$, shown in the bottom left inset) corresponds to a branch of period-doubled solutions that emanate from period-doubling (PD) bifurcation points indicated by diamond markers. The PD branch is computed and plotted as a gray line in Fig.~\ref{fig:3}(a), and its validity is confirmed using an event-driven direct numerical integration method, represented by gray stars. The period-doubled solutions (not shown) are not unilateral. Similar branches of period-doubled solutions have been reported in systems with bilinear elasticity~\cite{huang_dynamic_1989}.  Fig.~\ref{fig:3}~(b) shows the unilateral ratio in the same range of the forcing frequency. 

Although it is possible for the symmetric system to exhibit unilateral transmission, this characteristic occurs far from the system's resonance region. The response of the system has a very low amplitude and operating a mechanical system off-resonance is inefficient and often unproductive in practice. We were not able to find near-resonance unilateral transmission in the symmetric system by changing the bilinear ratio $\beta$ (including hardening), damping ratio $\zeta$ or forcing amplitude $P$. It is possible to overcome this issue, however, by breaking the mirror symmetry of the system.

\subsection{Unilateral transmission in an asymmetric system}
\label{locus}
\begin{figure}
\begin{subfigure}[b]{.24\textwidth}
  \includegraphics[width=\textwidth]{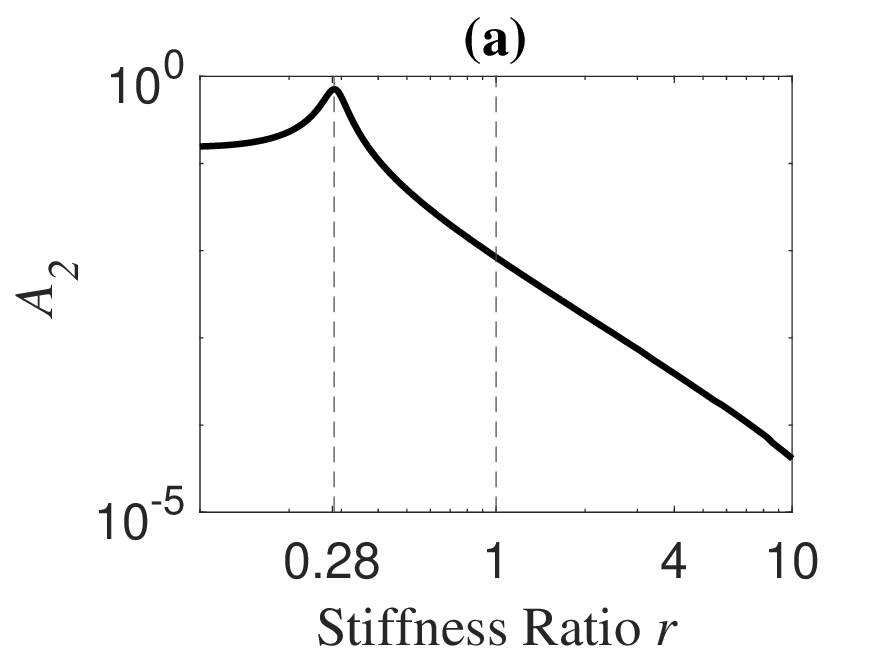}
\end{subfigure}
\hfill
\begin{subfigure}[b]{.24\textwidth}
      \includegraphics[width=\textwidth]{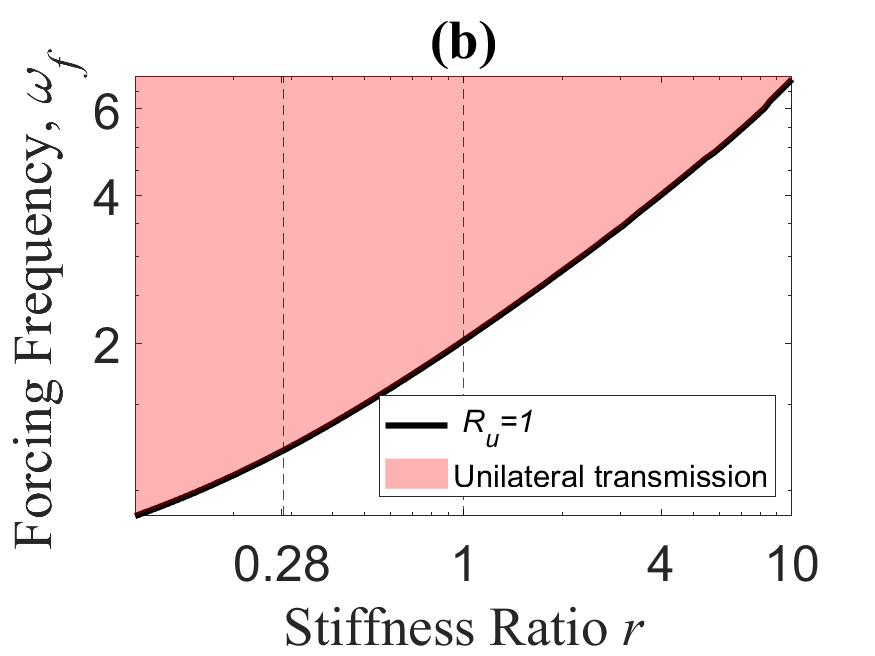}
\end{subfigure}

    \caption{Locus of the onset of unilateral transmission $R_u=1$ as a function of the stiffness ratio $r$ for (a) Output amplitude (b) Forcing frequency.}
    \label{fig:4}
\end{figure}

To find an asymmetric system that exhibits unilateral transmission near its primary resonances, we compute the locus of the onset of unilateral transmission, $R_u=1$, as a function of one of the symmetry-breaking parameters.
Fig.~\ref{fig:4}(a) shows the output amplitude (forward configuration) at the onset of unilateral transmission as a function of the stiffness ratio, $r$. We observe that lower values of the stiffness ratio (softer grounding stiffness for $M_1$) lead to higher output amplitudes at the onset of unilateral transmission. The maximum output occurs near $r=0.28$, which aligns the onset of unilateral transmission to the peak frequency.
Fig.~\ref{fig:4}(b) shows the variation of the forcing frequency along the same locus ($R_u=1$). The response at the forcing frequencies that lie above this locus, highlighted by a red background, exhibit unilateral transmission.

\begin{figure}
  \includegraphics[width=.5\textwidth]{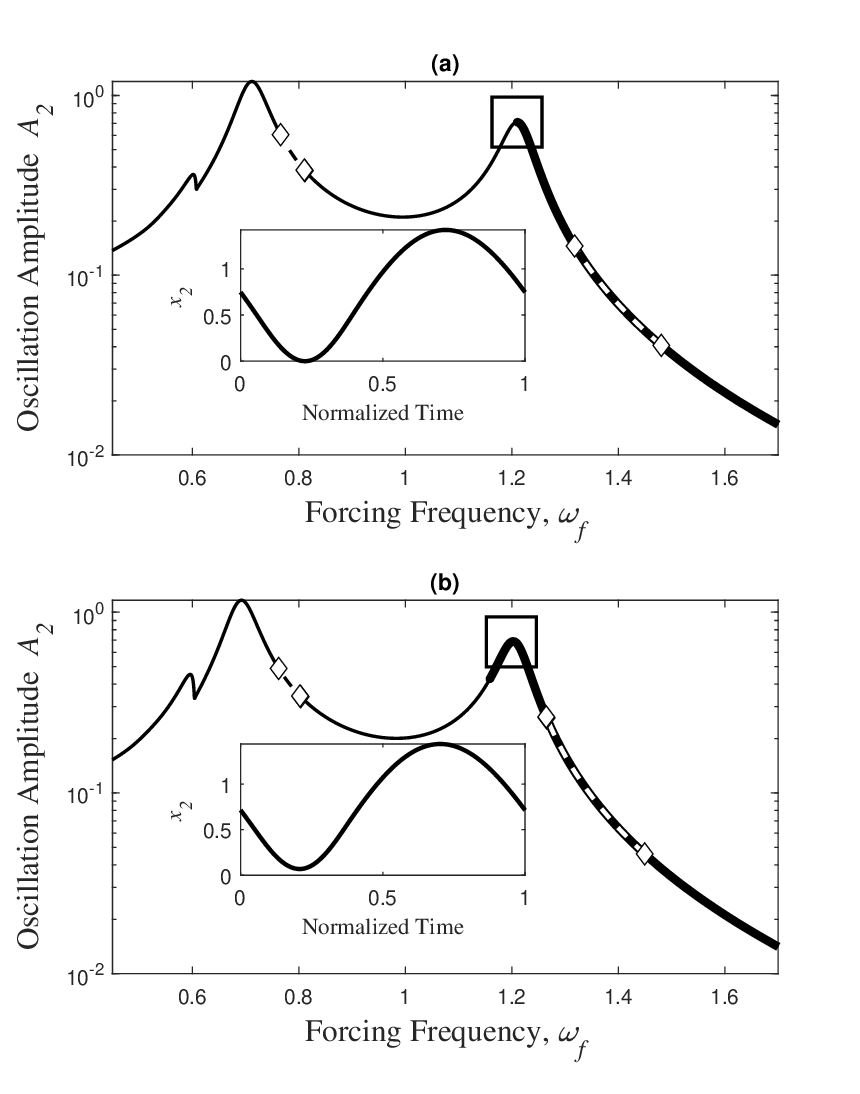}
  \centering
\caption{(a)~Frequency response of the asymmetric system $r=0.28$ for the onset of unilateral transmission at the peak frequency
(b)~Frequency response of the asymmetric system $r=0.25$ for the onset of unilateral transmission before the peak frequency}
\label{fig:5}      
\end{figure}

Fig.~\ref{fig:5}(a) shows the frequency response function of the asymmetric system with $r=0.28$. The inset shows the output displacement at the resonance peak indicated by the square over one forcing period. This point corresponds to the onset of unilateral transmission, which is characterized by the minimum value of the time-domain response {\it grazing} the static equilibrium position at zero. 

Because the onset of unilateral transmission occurs precisely at the second peak frequency, $R_u>1$ on one side of the peak and $R_u<1$ on the other side. Fig.~\ref{fig:4} suggests that selecting a lower value for the stiffness ratio ($r<0.28$) can shift the onset of unilateral transmission to a lower forcing frequency such that unilateral transmission covers the peak frequency on both sides.
Fig.~\ref{fig:5}(b) shows the frequency response function of the system for $r=0.25$. As expected, the region of stable unilateral response covers a wider frequency range around the peak frequency. The value of the unilateral ratio is $10\%$ larger than 1 at the peak ($R_u=1.1$). We also note that the onset of unilateral transmission has a strong dependence on the stiffness ratio, especially near the peak frequency.

\begin{figure}
  \includegraphics[width=.5\textwidth]{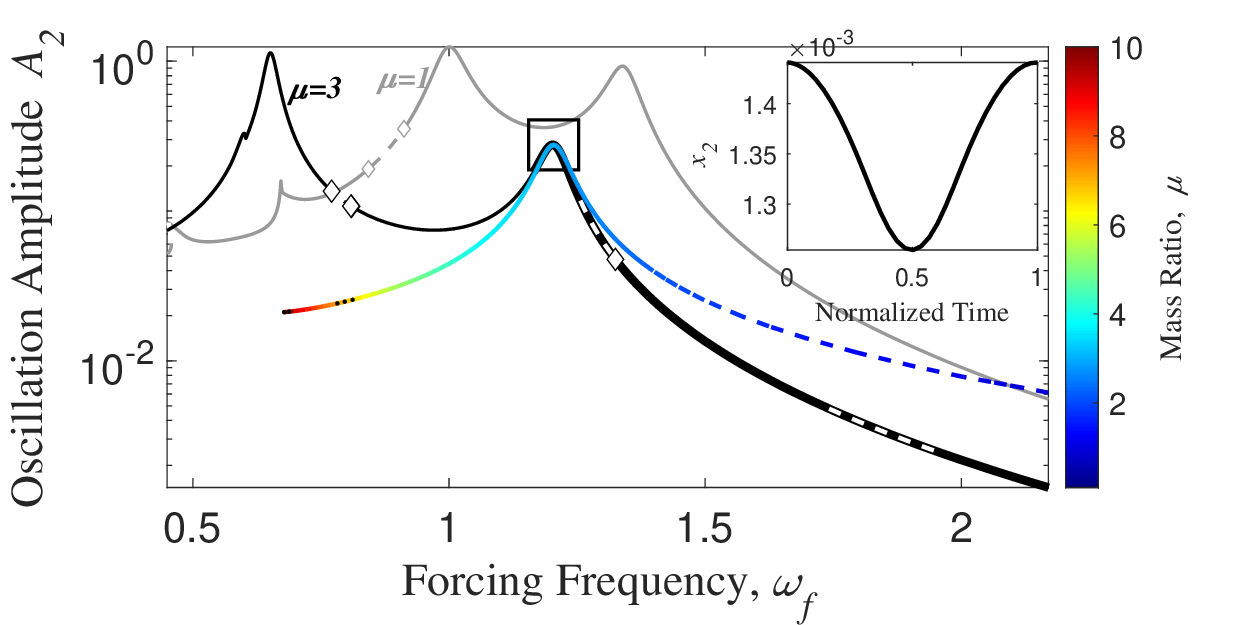}
  \centering
\caption{Frequency response of the system with $r=1$ and $\mu=3$. Thick lines indicate response that exhibits unilateral transmission and dashed lines indicate unstable response. The colored contoured line shows the locus of $R_u=1.1$. }
\label{fig:6}      
\end{figure}

We followed the same procedure for finding unilateral transmission occurring near a resonance of the system, this time using the other symmetry-breaking parameter, $\mu$, the mass ratio. To ensure that the region of unilateral transmission covers both sides of the peak frequency, we set $R_u=1.1$ based on the observations from Fig.~\ref{fig:5}(b).
Fig.~\ref{fig:6} shows the frequency response function of the system with $r=1$ and $\mu=3$.  The dashed contoured line shows the locus of $R_u=1.1$ and the color bar indicates the value of the mass ratio $\mu$ along this locus. Comparing Figures~\ref{fig:6} and~\ref{fig:5}(b), we observe that both $r$ and $\mu$ can result in unilateral transmission near the second primary resonance of the system.

\subsection{Influence of the bilinear ratio}
\label{beta}
The bilinear stiffness is the only source of nonlinearity in our system. To investigate the influence of bilinearity on unilateral transmission, we focus on the frequency range $1.16 < \omega_f < 2.2$, where unilateral transmission occurs for the system with $r=1$ and $\mu=3$; recall Fig.~\ref{fig:6}. This range is discretized into several frequencies. At each frequency, we compute the variation of the unilateral ratio ($R_u$) as a function of the bilinear ratio ($\beta$).

\begin{figure}
\begin{subfigure}[b]{.5\textwidth}
      \includegraphics[width=\textwidth]{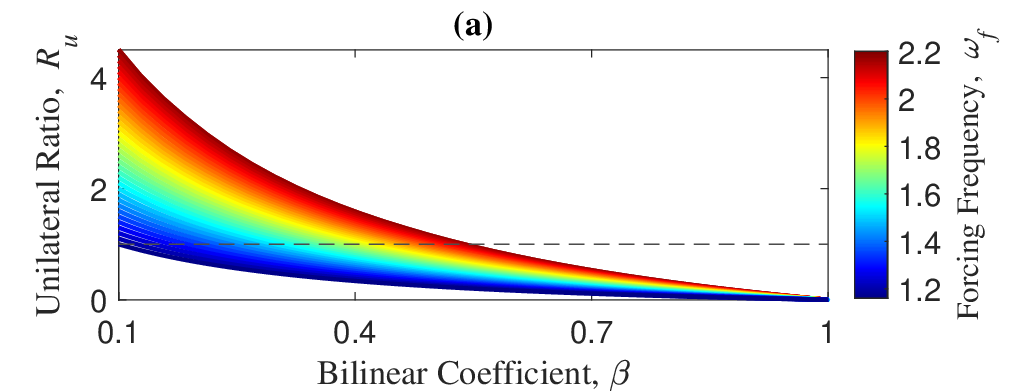}
\end{subfigure}
\hfill
\begin{subfigure}[b]{.5\textwidth}
      \includegraphics[width=\textwidth]{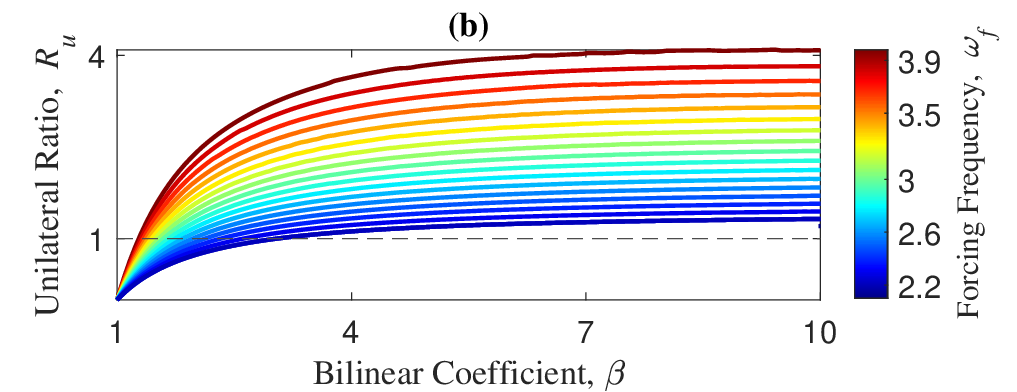}
\end{subfigure}

    \caption{Influence of bilinear ratio $\beta$ on unilateral transmission for $r=1$, $\mu=3$, $\zeta=0.03$, and $F=0.15$. (a) Softening bilinearity, $\beta<1$, (b) hardening bilinearity, $\beta>1$.}
    \label{fig:7}
\end{figure}

Fig.~\ref{fig:7}(a) shows a contour plot of $R_u$ as a function of $\beta$ for the case of softening bilinearity ($\beta < 1$). As $\beta$ increases, the frequency range with $R_u > 1$ becomes smaller, as indicated by the shrinking color range above the $R_u = 1$ line. As $\beta$ approaches 1, the system becomes linear and the unilateral ratio approaches zero for all forcing frequencies. As expected, all the curves converge at $(\beta,R_u)=(1,0)$. 
Fig.~\ref{fig:7}(b) shows a similar behavior for a hardening bilinear spring ($\beta > 1$). As the bilinear ratio (degree of nonlinearity) increases, there is an increasingly larger range of forcing frequencies over which unilateral transmission occurs.

\subsection{Nonreciprocal Unilateral Transmission}
\label{nonreciprocity}

We obtained near-resonance unilateral transmission in the system with broken mirror symmetry; Section~\ref{locus}. Because the system is nonlinear and asymmetric, it is therefore possible to find parameters that lead to nonreciprocal response. This property holds even though the response of the system is independent of the forcing amplitude in the parameter range that we consider in this work. Here, we investigate the possibility of nonreciprocal unilateral transmission of vibrations for the 2DoF system.

\begin{figure}[h!]
\begin{subfigure}{.5\textwidth}
  \includegraphics[width=\textwidth]{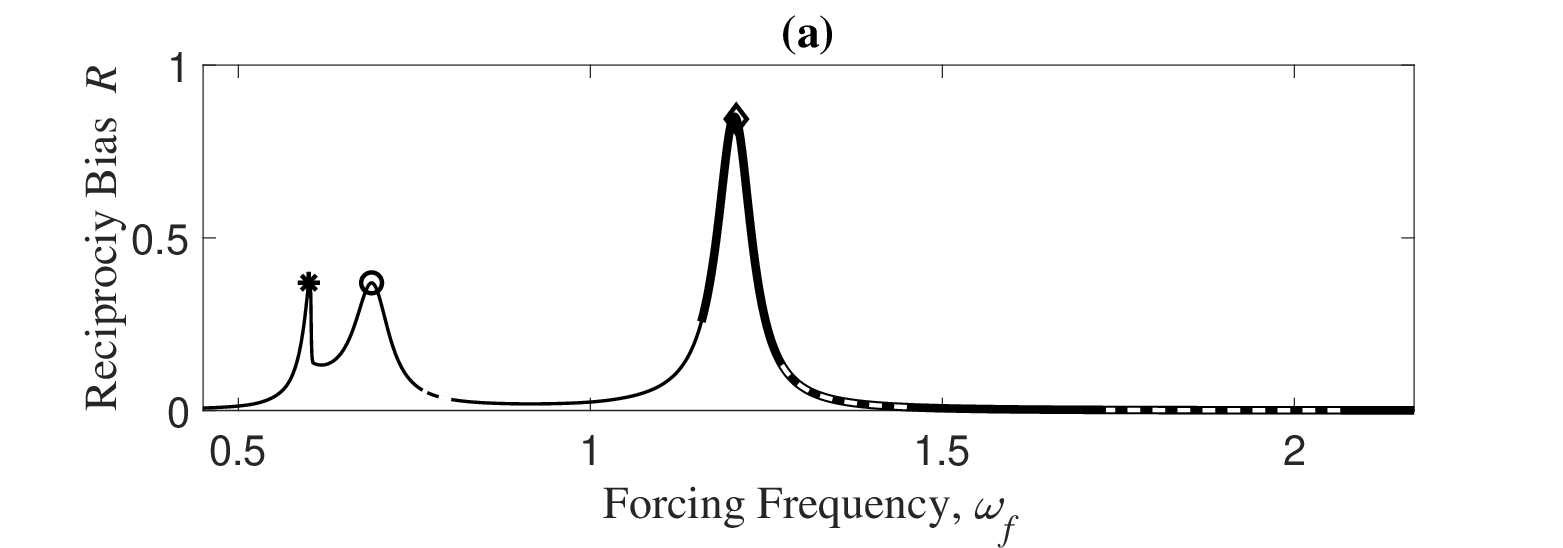}
\end{subfigure}

\begin{subfigure}[b]{.16\textwidth}
\includegraphics[width=\textwidth]{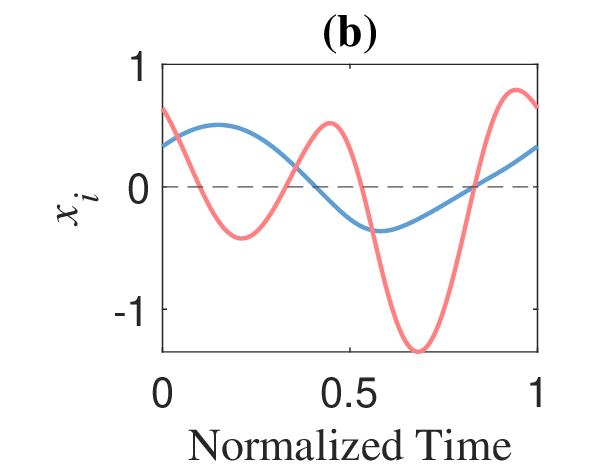}
\end{subfigure}
\begin{subfigure}[b]{.16\textwidth}
\includegraphics[width=\textwidth]{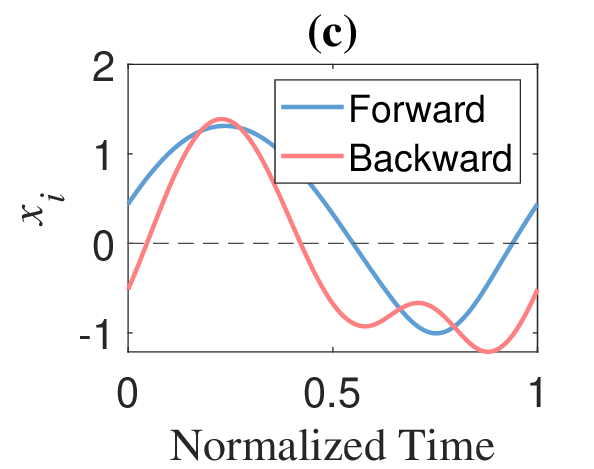}
\end{subfigure}
\hfill
\begin{subfigure}[b]{.16\textwidth}
\includegraphics[width=\textwidth]{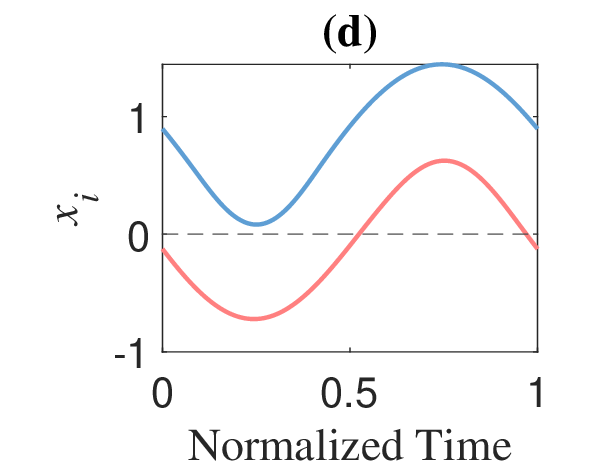}
\end{subfigure}
    \caption{
    (a) Reciprocity bias for the system with $r=0.25$ and $\mu=1$. Thick lines indicate response that exhibits unilateral transmission and dashed lines indicate unstable response in either the forward or backward configuration. (b) to (d) show the time-domain output displacement at the points indicated by the star, circle and diamond markers, respectively.}
  
\label{fig:8}
\end{figure}

Fig.~\ref{fig:8}(a) shows the reciprocity bias, $R$, for the system with $(\mu,r)=(1,0.25)$. As expected, the system exhibits nonreciprocal response. Figures~\ref{fig:8}(b) to (d) show the time-domain response of the system at the local maxima of $R$, indicated by a star, circle and diamond respectively. In both panels (b) and (c), the response of the forward configuration is primarily harmonic while that of the backward configuration has significant contributions form the second harmonic. Thus, we observe harmonic generation only in one direction. At the global maximum of the reciprocity bias, panel (c), we observe unilateral nonreciprocal transmission. This corresponds to the second primary resonance of the system where the masses move out of phase, thereby engaging the bilinear spring significantly. 

\begin{figure}
\begin{subfigure}{.5\textwidth}
\includegraphics[width=1\textwidth]{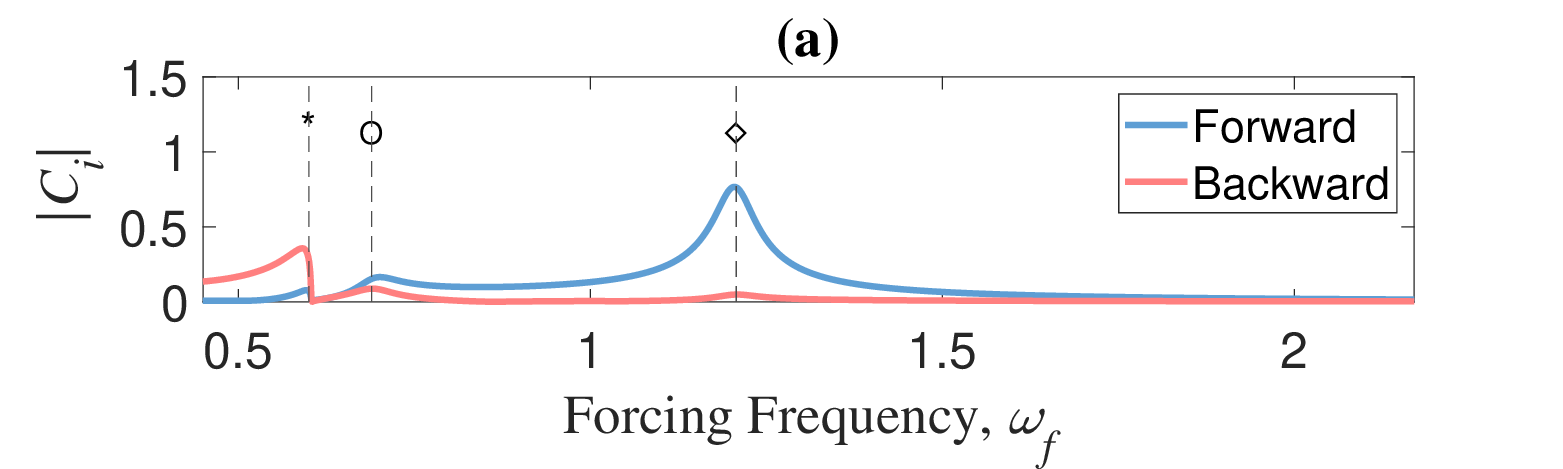}
\end{subfigure}
\begin{subfigure}{.5\textwidth}
\includegraphics[width=1\textwidth]{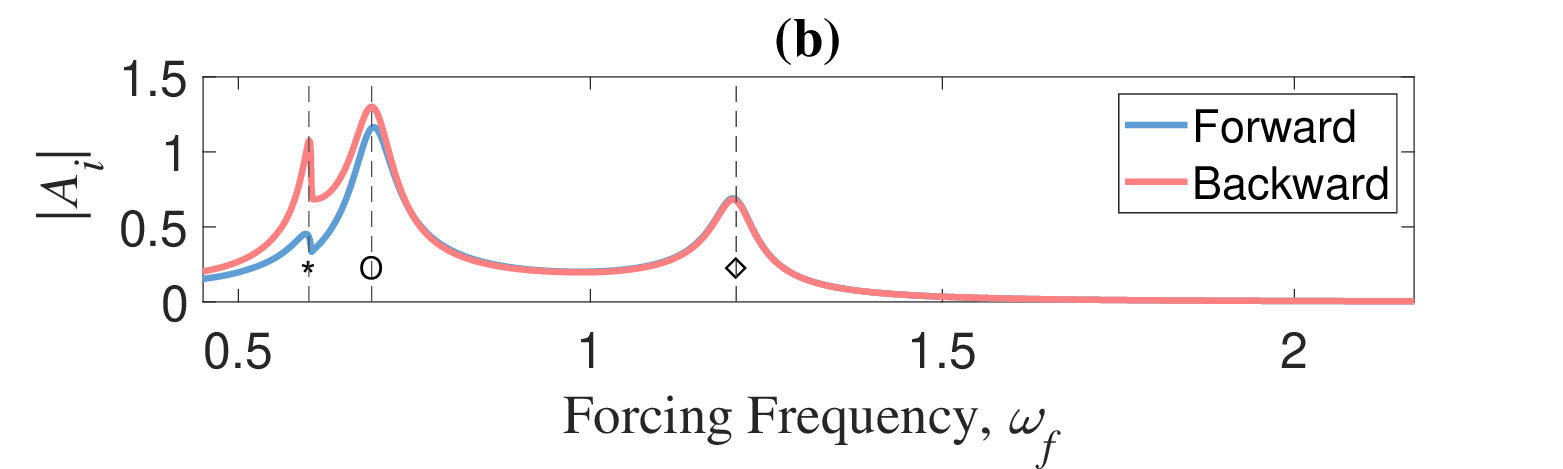}
\end{subfigure}
\begin{subfigure}{.5\textwidth}
\includegraphics[width=1\textwidth]{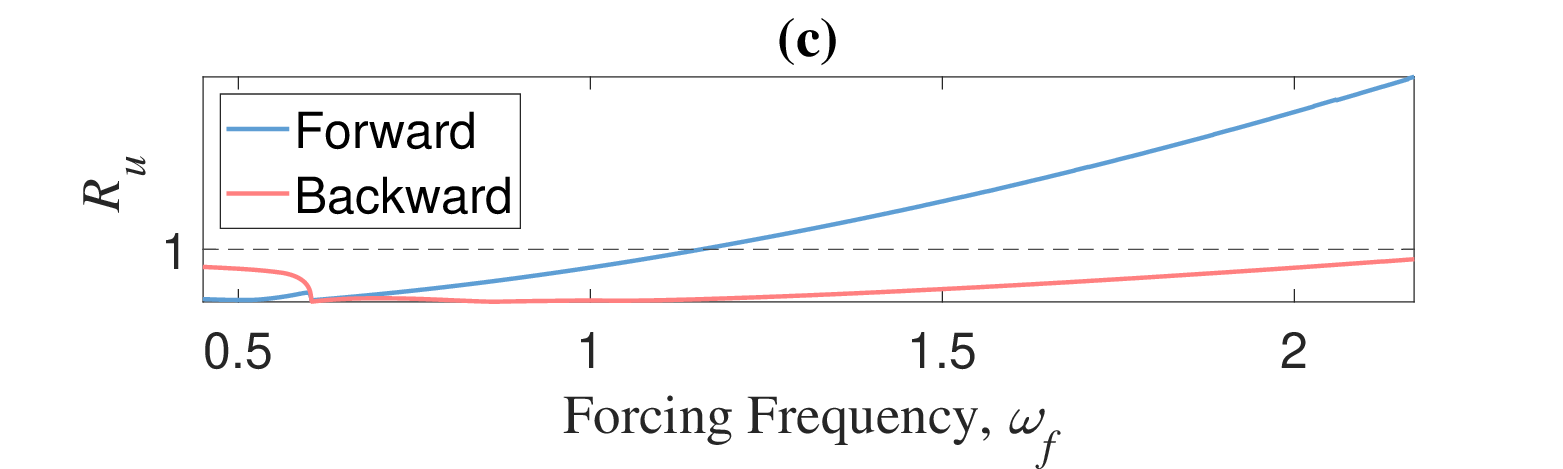}
\end{subfigure}
    \caption{(a) DC value of the response for the system with $r=0.25$ and $\mu=1$ (b) AC value of the response
(c) Unilateral ratio for the forward and backward configurations}
\label{fig:9}
\end{figure}

Figures~\ref{fig:9}(a) and (b) show the DC and AC components of the output displacement, respectively. The DC component is responsible for enabling unilateral transmission. At the diamond marker in Fig.~\ref{fig:9}(a,b), there is a significant difference between the DC values in the forward and backward configurations, while the AC values are similar for the two configurations. Figures~\ref{fig:9}(c) shows the variation of the unilateral ratio, $R_u$, over the same interval of the forcing frequency. As expected, the onset of unilateral transmission in the forward configuration is accompanied by the increase in the DC value of the output displacement.  

\section{Periodic Structure with Bilinearly Coupled Units}
\label{periodic}
\begin{figure*}
  \includegraphics[width=1\textwidth]{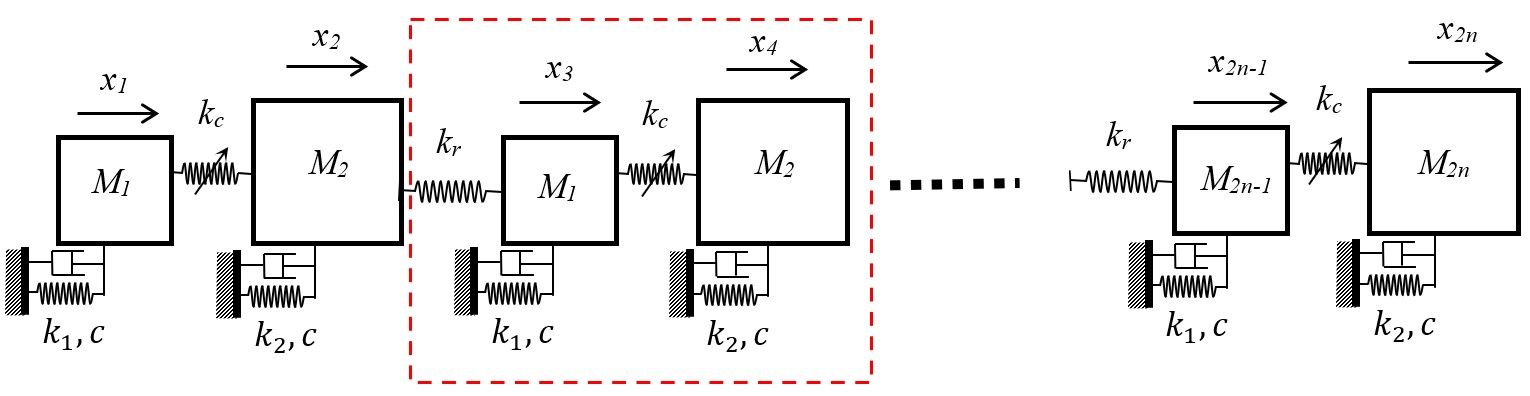}
  \centering
\caption{Periodic structure with a bilinearly coupled unit cell }
\label{fig:10}      
\end{figure*}

Fig.~\ref{fig:10} shows a periodic lattice composed of $n$ unit cells. The unit cells are the same as the 2DoF system in Fig.~\ref{fig:1}, with adjacent units coupled to each other with a linear spring of non-dimensional stiffness $k_r$. In Section \ref{sec:1}, we found sets of system parameters that lead to stable unilateral transmission within a unit cell near a peak frequency. Building on this design, we investigate unilateral vibration transmission in the end-to-end steady-state response of the periodic structure in Fig.~\ref{fig:10} to external harmonic excitation. In particular, we highlight the influence of two key parameters of the periodic system on its vibration transmission characteristics: the number of units and energy dissipation.

For the results presented in this section, each unit cell featured an asymmetry in the mass ratio only, $(\mu,r)=(3,1)$, and a softening bilinear ratio of $\beta=0.1$. This is similar to the case studied in  Sec.~\ref{locus}, Fig.~\ref{fig:6}. The non-dimensional constant of the spring connecting adjacent unit cells is $k_r=5$, resulting in well separated band of resonances in the finite periodic structure.

\subsection{Influence of the number of units}
We investigate unilateral transmission in lattices of 2, 4, 8, and 16 unit cells. Extension of the results to infinitely long systems requires a different methodology and lies beyond the scope of the present work. 

\begin{figure*}[h!]
\centering
\begin{subfigure}{1\textwidth}
  \includegraphics[width=\textwidth]{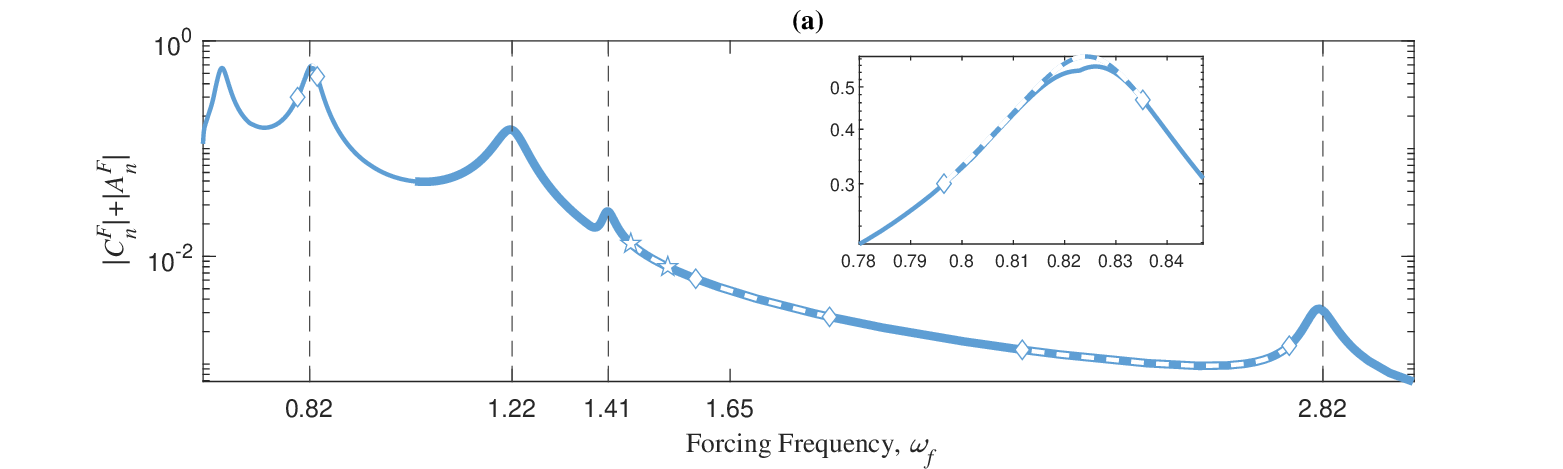}
\end{subfigure}

\begin{subfigure}{1\textwidth}
  \includegraphics[width=\textwidth]{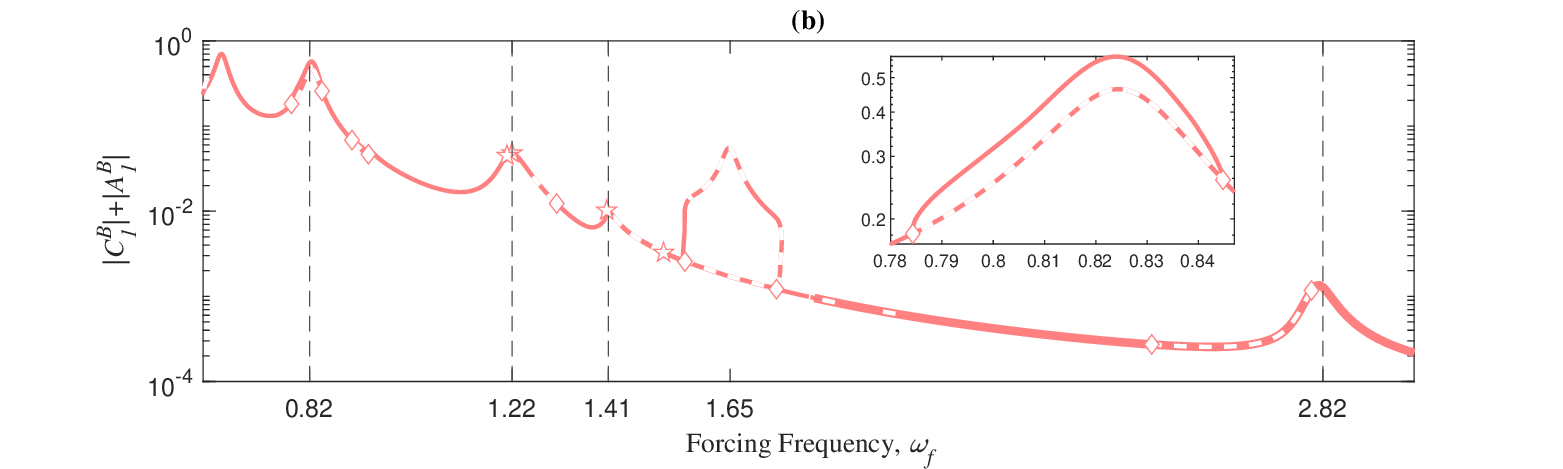}
\end{subfigure}
\hfill
\begin{subfigure}{.24\textwidth}
  \includegraphics[width=\textwidth]{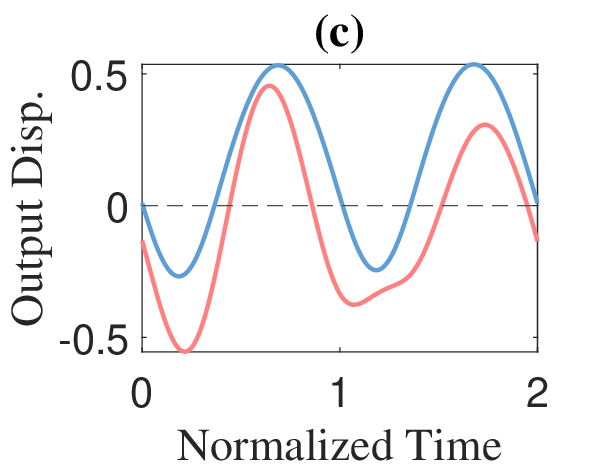}
\end{subfigure}
\hfill
\begin{subfigure}{.24\textwidth}
  \includegraphics[width=\textwidth]{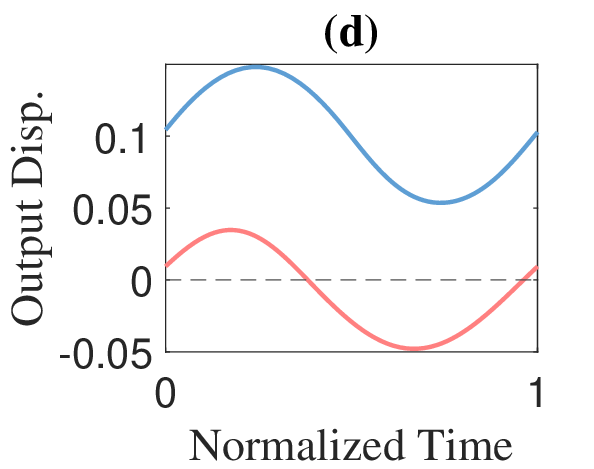}
\end{subfigure}
\hfill
\begin{subfigure}{.24\textwidth}
  \includegraphics[width=\textwidth]{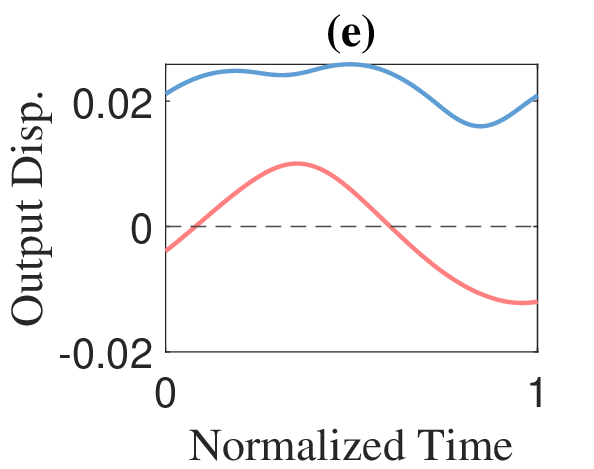}
\end{subfigure}
\hfill
\begin{subfigure}{.24\textwidth}
  \includegraphics[width=\textwidth]{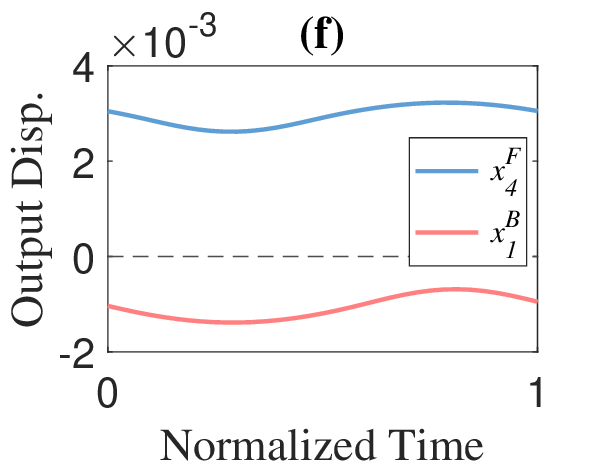}
\end{subfigure}
\caption{Frequency response of the system for $n=2$, $\mu=3$ and $r=1$ for (a) forward configuration and (b) backward configuration. Diamonds and pentagrams denote period-doubling and torus bifurcations, respectively. Time-domain response of the system for the forward and backward configurations at (c) $\omega_f=0.82$ (d) $\omega_f=1.22$, (e) $\omega_f=1.41$ and (f) $\omega_f=2.82$ indicated by vertical lines. In panel (e), $x_1^B(t)$ is computed using direct numerical integration.}
\label{fig:11}
\end{figure*}

Fig.~\ref{fig:11}(a) and (b) illustrate the response of the periodic structure with two unit cells ($n=2$, 4DoF in total) for the forward and backward configurations, respectively. As in previous sections, unilateral response is indicated in the frequency response function by thicker lines. Four distinct frequency peaks are identified by vertical lines, and their corresponding time response for the forward and backward configurations are shown in Fig.~\ref{fig:11}(c)-(f). 

Near the second resonant peak ($\omega_f=0.82$), the system undergoes a PD bifurcation in both forward and backward configurations, as seen in Fig.~\ref{fig:11}~(c). The corresponding branches of period-doubled solutions are shown in the insets in panels (a) and (b). At the two subsequent peaks, shown in panels (d) and (e), the system exhibits nonreciprocal unilateral transmission: the response is unilateral only in the forward configuration. At $\omega_f=1.41$, panel (e), the backward configuration has undergone a torus bifurcation, resulting in the quasi-periodic response. At $\omega_f=2.82$, panel (f), the response of the system is nonreciprocal but unilateral for both forward and backward configurations.

We note that the system exhibits several other bifurcations. The bifurcation structures in the forward and backward configurations are different from each other, as indicated by the markers in Fig.~\ref{fig:11}(a)-(b). As a case in point, there is a pair of period-doubling bifurcations occurring near $\omega_f\approx1.65$. The branch of the period-doubled solution is shown for the backward configuration in panel Fig.~\ref{fig:11}(b), which undergoes secondary bifurcations. We have found period-doubling bifurcations to generally increase the amplitude of motion especially when they occur away from a primary resonance. 

\begin{figure}[h!]
\centering
\begin{subfigure}{.5\textwidth}
  \includegraphics[width=\textwidth]{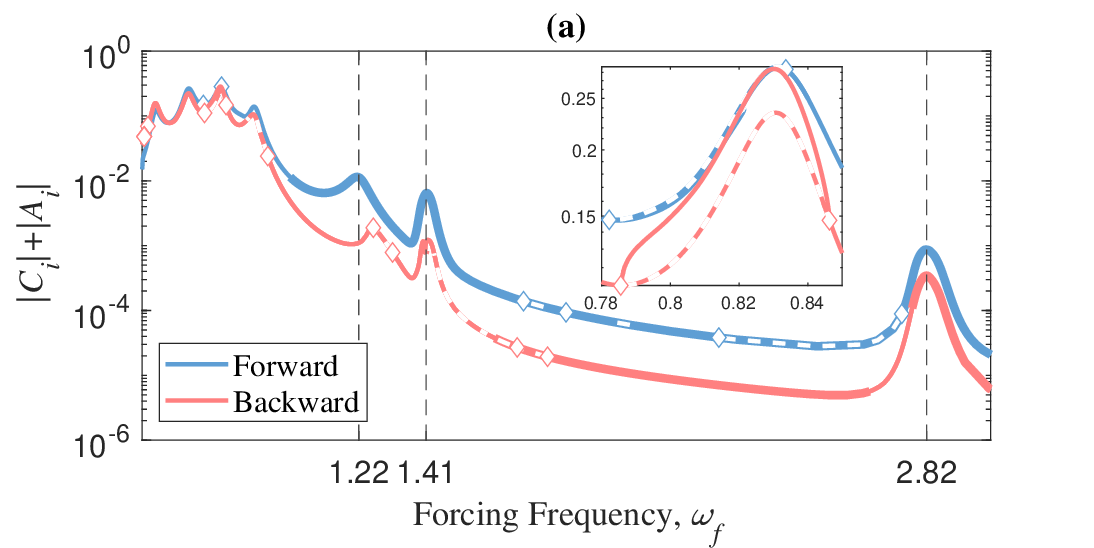}
\end{subfigure}

\begin{subfigure}{.5\textwidth}
  \includegraphics[width=\textwidth]{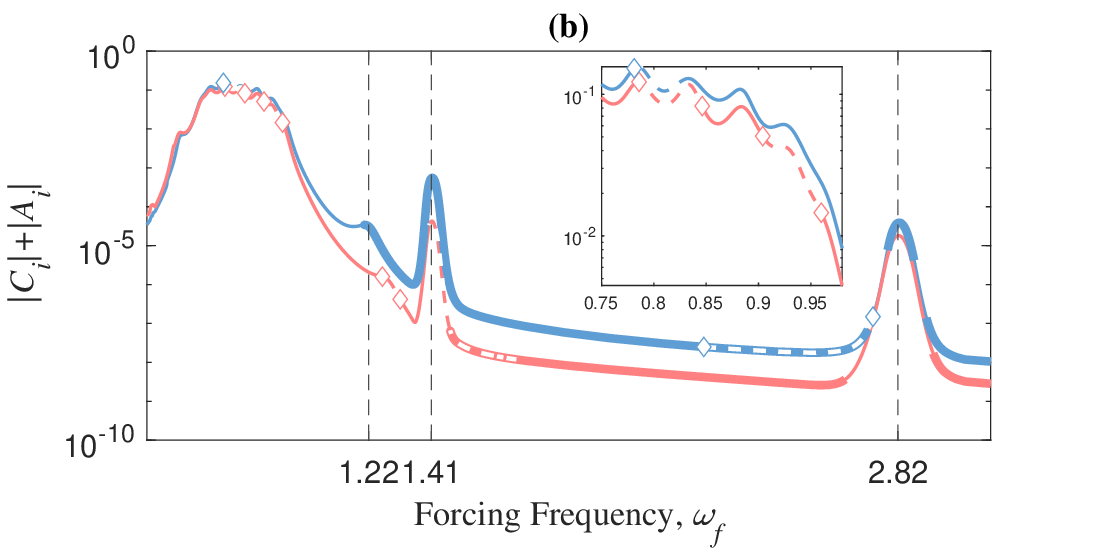}
\end{subfigure}

\begin{subfigure}{.5\textwidth}
  \includegraphics[width=\textwidth]{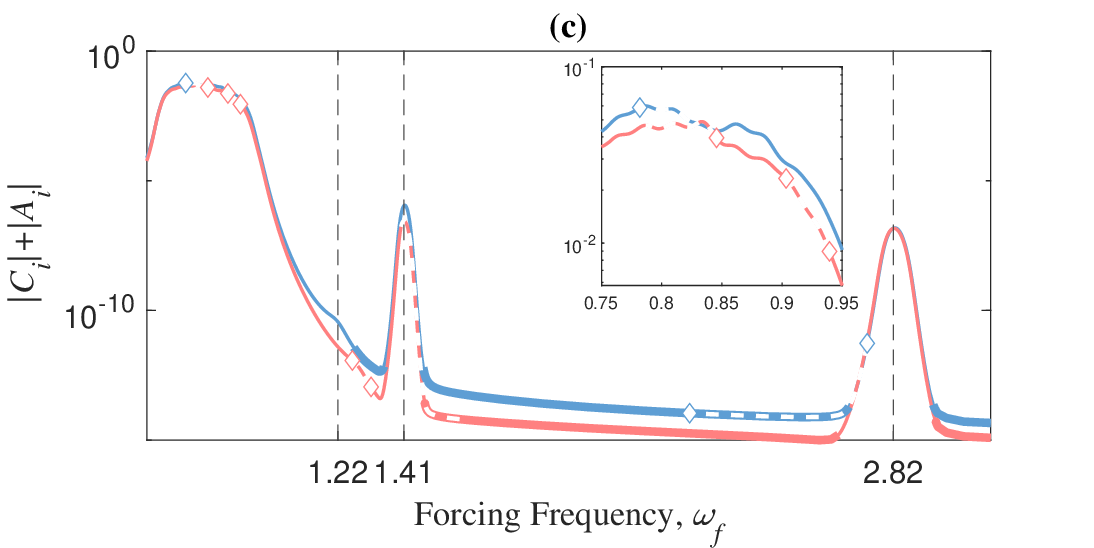}
\end{subfigure}
\hfill

\caption{Frequency response curves of the forward and backward configurations for $\mu=3$, $r=1$, and $\zeta=0.03$ for (a) $n=4$, (b) $n=8$, and (c) $n=16$. Diamond markers denote period-doubling bifurcations.}
\label{fig:12}
\end{figure}

Figures~\ref{fig:12}(a)-(c) display the response of the periodic structure for $n=4$, $n=8$, and $n=16$, respectively. The insets highlight the frequency response curves for the in-phase modes (acoustic band). All systems undergo period-doubling bifurcations near $\omega_f\approx0.82$. As an example, the inset in Fig.~\ref{fig:12}(a) shows the ensuing branch of period-double solutions for the system with $n=4$. The periodic-doubled solutions are not reciprocal (time-domain response not shown). 

The peak near $\omega_f\approx2.82$ corresponds to the out-of-phase modes (optical band). The $n$ modes appear as one damped peak because of the modal overlap caused by damping; see~\cite{kogani_nonreciprocal_2024} for a similar situation. As expected, the response amplitude at the output decreases with increasing the number of unit cells because of damping. This is more pronounced in the optical band because adjacent masses move out of phase with each other in this frequency range. 

For $n=4$, shown in Fig.\ref{fig:12}(a), the system exhibits unilateral transmission at $\omega_f=1.22$, $\omega_f=1.41$ and $\omega_f=2.82$. The system with $n=2$ exhibited a similar behavior (Fig.\ref{fig:11}(a) and (b)). The first two (lower) frequencies correspond to nonreciprocal unilateral transmission, where the response is unilateral in the forward configuration, while the third (higher) frequency exhibits unilateral transmission in both the forward and backward configurations.

For $n=8$, Fig.~\ref{fig:12}(b), the backward configuration no longer exhibits unilateral transmission at any of the three forcing frequencies in Fig.~\ref{fig:12}(a). However, the response of the forward configuration remains unilateral at these frequencies. We observe that the response of the system is unilateral over a shorter range of frequencies as a result of increasing the number of units. 

Fig.~\ref{fig:12}(c) shows the response of the system with $n=16$, where unilateral transmission no longer occurs near any of the peaks. This is attributed to the increased energy dissipation in the longer periodic structure.

\begin{figure}[h!]
\centering
\begin{subfigure}{.5\textwidth}
  \includegraphics[width=\textwidth]{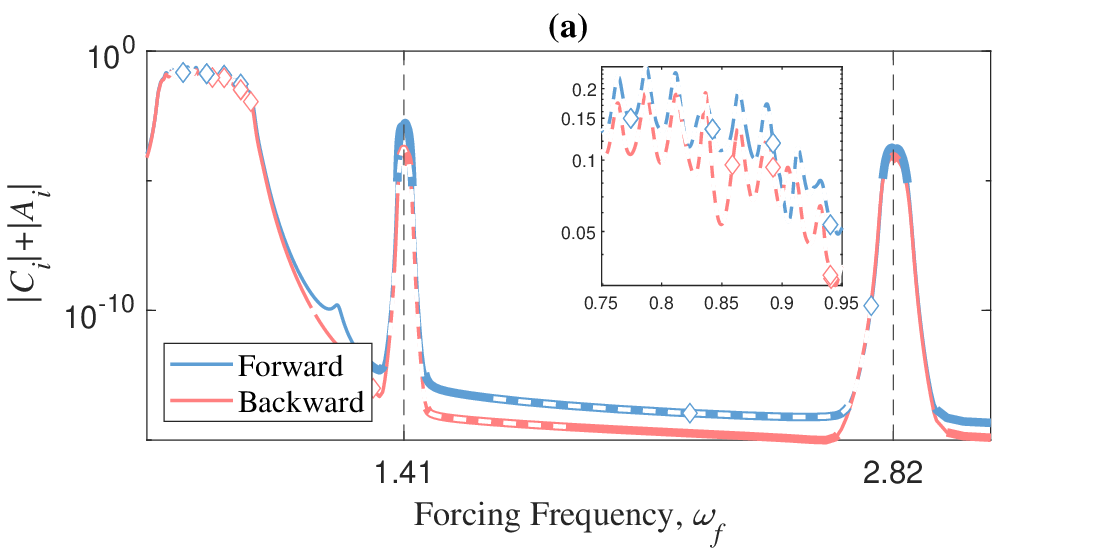}
\end{subfigure}
\begin{subfigure}[b]{.21\textwidth}
\includegraphics[width=\textwidth]{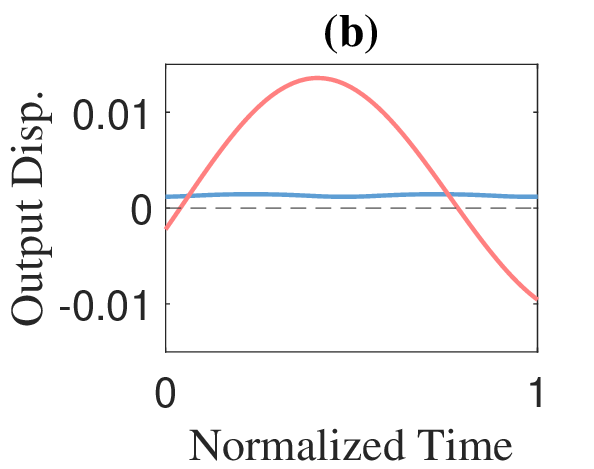}
\end{subfigure}
\hfill
\begin{subfigure}[b]{.21\textwidth}
\includegraphics[width=\textwidth]{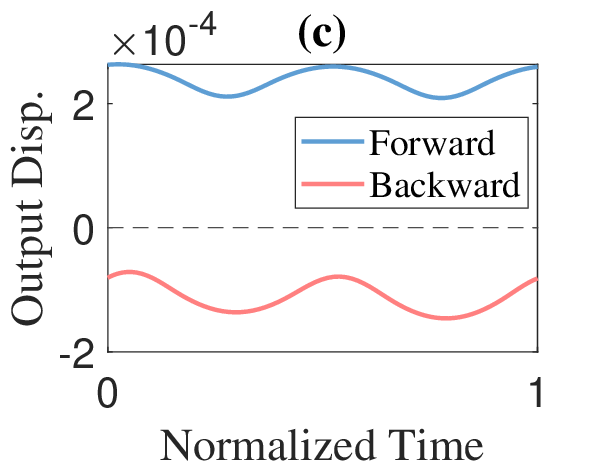}
\end{subfigure}

\hfill

\caption{ (a) Frequency response curves of the forward and backward configurations for $n=16$, $\mu=3$, $r=1$, and $\zeta=0.01$. Time-domain response of the system for the forward and backward configurations at (b) $\omega_f=1.41$ and (c) $\omega_f=2.82$. In panel (b), $x_1^B(t)$ is computed using direct numerical integration.}
\label{fig:13}
\end{figure}

\subsection{Influence of damping}

mber of unit cells increases, vibrations are increasingly more suppressed as they propagate through the structure. Therefore, the output displacement in periodic structures with a large number of unit cells is significantly damped. This energy dissipation decreases the influence of nonlinearity, which is required for unilateral transmission to occur. Fig.~\ref{fig:12}(c) shows that the system with $n=16$ no longer exhibits unilateral transmission near peak frequencies.

To further clarify the role of damping in suppressing unilateral transmission, Fig.~\ref{fig:13}(a) shows the response of the system with $n=16$ at a lower damping ratio of $\zeta=0.01$; {\it cf.} Fig.~\ref{fig:12}(c). Unilateral transmission is restored in the system with lower damping at both peaks. At $\omega_f=1.41$, panel (b), only the forward configuration exhibits unilateral transmission, while at $\omega_f=2.82$, panel (c), the response is unilateral for both configurations. The response of the backward configuration is quasiperiodic at $\omega_f=1.41$. 
We note that the amplitude of the output displacements remains very small; vibrations are transmitted through 32 damped units to reach the last unit, in this case.

Another feature of the response of the long structure with low damping is instability of the response, specifically near the primary in-phase resonances (acoustic band); see the inset in Fig.~\ref{fig:13}. Period-doubling bifurcations appear to be the main mechanism for loss of stability in the acoustic band, while the instability at $\omega_f=1.41$ is caused by a torus bifurcation. Investigating the detailed structure of these secondary bifurcations falls beyond the scope of the present work and warrants a separate analysis.

\clearpage
\section{Conclusions}
\label{conclusion}

We investigated the phenomenon of unilateral transmission, where incoming harmonic waves attain a single sign (pure tension or compression) upon transmission through the system. We demonstrated this behavior in a system with bilinear elasticity, exhibiting different effective stiffness in compression and tension. We analyzed the steady-state response of a two-degree-of-freedom (2DoF) system with bilinear elasticity subject to external harmonic excitation using numerical continuation. We found that unilateral transmission can occur in a system with mirror symmetry (identical oscillators, coupled bilinearly), but this occurred away from resonant frequencies.

We demonstrated that breaking the mirror symmetry of the system allows unilateral transmission to occur near the resonances of the system. Specifically, both the stiffness ratio and the mass ratio of the system can be adjusted to enable near-resonance unilateral transmission. The introduction of asymmetry also resulted in nonreciprocal dynamics, which we explored in the context of unilateral transmission. We reported on response regimes that lead to nonreciprocal unilateral transmission: when unilateral transmission occurs in one direction but not the other. We also found nonreciprocal dynamics that is characterized by harmonic generation only in one direction. 

Building on these findings, we extended our analysis to vibration transmission in a periodic structure that has the bilinear 2DoF system as its unit cell. We investigated the influence of the number of unit cells and energy dissipation on unilateral transmission. We demonstrated that stable nonreciprocal unilateral transmission can occur near the primary and internal resonances of the system, including a regime in which both the forward and backward configurations exhibit unilateral transmission (one in pure compression, the other in pure tension). We showed that as the number of unit cells increases from 2 to 16, energy dissipation can suppress unilateral transmission. 

We hope that these findings on unilateral transmission open new avenues in the operation of vibration systems, providing insights for the design and optimization of such systems in engineering applications and devices.

\section*{Acknowledgments}
We acknowledge financial support from the Natural Sciences and Engineering Research Council of Canada through the Discovery Grant program. A.K. acknowledges additional support from Concordia University. 

\nocite{*}





\bibliographystyle{unsrt} 
\bibliography{mypaper.bib}


\section*{Appendix A: Non-dimensional Equations of Motion}
\label{appA}

The governing equations for the system in Fig.~\ref{fig:1} can be written as:
\begin{equation}
\begin{aligned}
\label{EOMraw}
M_1\bar{x}''_{1}+k_c(\bar{x}_1-\bar{x}_2)+k_1\bar{x}_1+c\bar{x}'_1=f_1\cos{\bar{\omega}_ft} \\
M_2\bar{x}''_2+k_c(\bar{x}_2-\bar{x}_1)+k_2\bar{x}_2+c\bar{x}'_2=f_2\cos{\bar{\omega}_ft}\\
\end{aligned}
\end{equation}
where $k_c$ is the coupling bilinear stiffness, $k_1$ and $k_2$ linear grounding stiffness for $M_1$ and $M_2$, and $c$ is the linear viscous damping connecting each mass to the ground. We divide the equations by $k_1$ and introduce the non-dimensional parameters $\tau=\omega_0t$, $\omega_0^2=k_1/M_1$, $\omega_f=\bar{\omega}_f/\omega_0$ to obtain 
\begin{equation}
\begin{aligned}
    \frac{M_1\omega_0^2}{k_1}\ddot{\bar{x}}_1 &+ \frac{k_c}{k_1}(\bar{x}_1-\bar{x}_2) + \bar{x}_1 + 2\zeta \dot{\bar{x}}_1 
    = \frac{f_1}{k_1}\cos{\omega_f\tau}, \\
    \frac{M_2\omega_0^2}{k_1}\ddot{\bar{x}}_2 &+ \frac{k_c}{k_1}(\bar{x}_2-\bar{x}_1) + \frac{k_2}{k_1}\bar{x}_2 + 2\zeta \dot{\bar{x}}_2 
    = \frac{f_2}{k_1}\cos{\omega_f\tau}.
\end{aligned}
\end{equation}
\\
where $\dot{x}=dx/d\tau=(dx/dt)/\omega_0$, $\ddot{x}=d^2x/d\tau^2=(d^2x/dt^2)/\omega_0^2$ and $\zeta=(c\omega_0)/(2k_1)$. 
We define the non-dimensional displacement and force as $x=\bar{x}/d$ and $F=f/(dk_1)$, where $d$ is a characteristic displacement of the system. This results in 
\begin{equation}
\label{eqNonD}
\begin{aligned}
\ddot{x}_1+K_c(x_1-x_2)+x_1+2\zeta\dot{x}_1=F_1\cos{\omega_f\tau} \\
\mu\ddot{x}_2+K_c(x_2-x_1)+rx_2+2\zeta\dot{x}_2=F_2\cos{\omega_f\tau}
\end{aligned}
\end{equation}
where $\mu=M_2/M_1$, $K_c=k_c/k_1$, and $r=k_2/k_1$. 
Eq.~(\ref{eqNonD}) is the non-dimensional form of Eq.~(\ref{EOMraw}). Eq.~(\ref{eqNonD}) is the same as Eq.~(\ref{govern}) in the main text, where we have replaced $\tau$ with $t$ for ease of reference.
\section*{Appendix B: Amplitude independency of bilinear systems without offset}
\label{app:B}
\renewcommand{\thefigure}{B\arabic{figure}} 
\setcounter{figure}{0} 
\label{appA}
In bilinear stiffness systems without an offset, the transfer function remains unaffected by amplitude variations, preserving a linear relationship with the input. Fig.~\ref{fig:B1} illustrates this observation by presenting the transfer function of the symmetric system under two distinct forcing amplitudes. 

\begin{figure}

  \includegraphics[width=.5\textwidth]{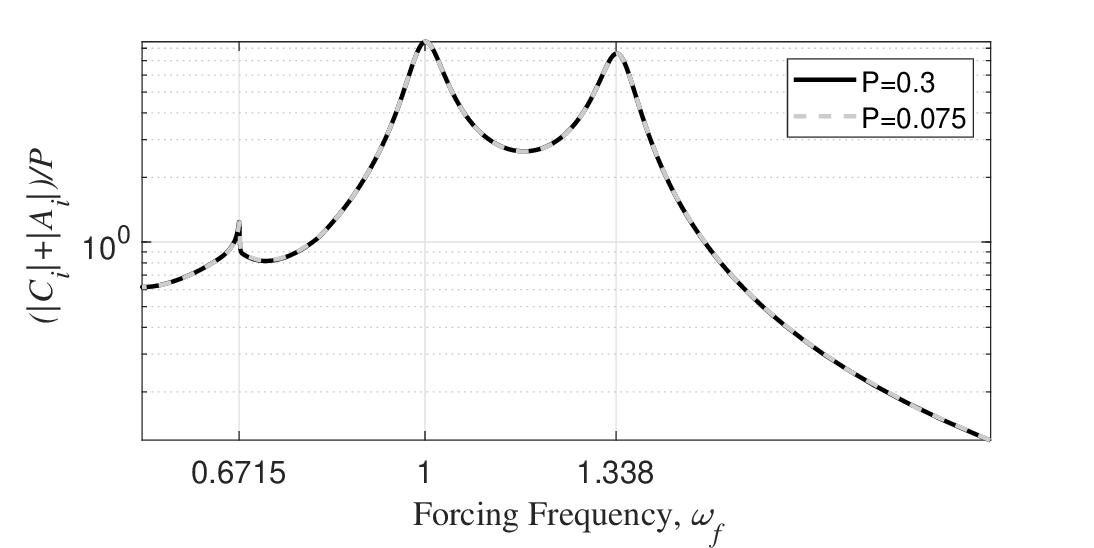}
  \centering
\caption{Frequency response function of the symmetric 2DoF system for $r=\mu=1$, $\zeta=0.03$, and $\beta=0.1$, at two forcing amplitudes. }
\label{fig:B1}      
\end{figure}

\end{document}